\documentclass[twocolumn,10pt]{article}
\usepackage{epsfig}
\makeatletter
\renewcommand{\section}{\@startsection
   {section}
   {1}
   {\z@}
   {-10mm}
   {2mm}
   {\normalfont\normalsize\bfseries}}
\renewcommand{\subsection}{\@startsection
   {subsection}
   {2}
   {\z@}
   {-5mm}
   {1mm}
   {\normalfont\normalsize\itshape}}
\makeatother

\begin{document}
\twocolumn[
{\Large \bf \textsf{On Seyfert spectra and variability}
\par}
\vspace*{2mm}
P.~Magdziarz
\par
\vspace*{2mm}
{\small Jagiellonian University, Astronomical Observatory, Cracow,
Poland\\
[1mm]}

{\it \small Received December, 1997; Accepted April, 1998} \par
\vspace{15mm}
]

\section*{Abstract}

{\small 
We discuss the observational phenomenology of Seyferts from the point of view
of the disk model, and argue that the spectral variability may be related
to geometrical changes of the cold matter which provides a source of
seed photons for Comptonization in a hot central disk.  One possible
configuration is a model with quasi-spherical accretion
in the central part of the disk, with variability determined by the dynamics of
the transition zone between the cold and hot disk.
The soft excess component appears as a driver of spectral variability and
contributes significantly to the source energetics. 
}

\section{Introduction}

The physical interpretation of the broad-band spectra of Seyferts nuclei is
still an open question, despite the fact that the actual radiation processes
responsible are well understood (e.g., Zdziarski et al.\ 1997).
One of the main current problems to solve is a mapping of the observed,
three component continuum emission, namely the big blue bump, the soft
excess, and the hard continuum, onto a physical picture of matter accreting
onto a black hole. Distinguishing between various geometries of AGN (e.g.,
Haardt 1997) is still an observational question on interpretation of the
broad-band spectrum and variability. 

\section{Variability}

It has been suggested that the complexity of AGN light curves is generally
related to the basic nonlinearity of the underlying physics (e.g., Vio et al.\
1992). Nonlinear variability has been clearly detected in high quality
optical observations of some bright objects (e.g., Vio et al.\ 1991,
3C~345; Longo et al.\ 1996, NGC~4151), however, searches for nonlinear
signatures have failed in most of the available X-ray data (e.g., Czerny and
Letho 1997).  This has been explained by the strong dependence of the analysis
methods on signal degradation due to observational noise (Leighly 1997). 
In fact, nonlinearity has been recently detected in some X-ray
observations with high signal to noise ratio on day-to-month time scales
relevant to energy reprocessing (Leighly and O'Brien 1997, 3C~390.9;
Boller et al.\ 1997, IRAS 13224-3809). 

The overall character of variability detected in active objects is
consistent with self-organized criticality models (SOC models;  Mineshige,
Ouchi and Nishimori 1994). This has been interpreted as a physical effect
of triggering a number of energy reservoirs by an instability in the disk.
This instability leads to an avalanche of accretion and in effect to
flares with substantial nonlinearity (i.e., internal correlation between
parameters of the flares).  The light curve then consists of a weakly
variable and relatively faint period before the triggering event, and
a bright strongly variable period with some kind of decay after the
triggering event. Recently, Leighly and O'Brien (1997; their Figure~1)
separated such a characteristic shape of the X-ray light curve in the case of
3C~390.9.  Some evidence for a similar pattern has also been detected in
optical observations of NGC~4151 and NGC~5548 (Cid Fernandes et al.\
1997). 

\section{Reprocessing and the source structure}

Turner et al.\ (1997) have suggested recently that the broad iron line in
Seyferts has a universal profile independent of type, leading to trouble for
unification schemes, and putting into question the relation between the iron
line and reflection component. Some Seyfert 2s show spectrum dominated by
reflection component which is, generally, explained by hiding the nucleus
by a large scale torus (e.g., Matt 1997). However, in the case of the
reflection dominated spectrum of MCG~-6-30-15 (type 1 Seyfert) both the
iron line and the reflection component are probably concentrated inside
about 10 Schwarzschild radii (Maciolek-Nied\'zwiecki et al.\ 1998). This,
together with the universal shape of the iron line, argues for a
quasi-spherical accretion in the central part of the disk.  The
structure of the central disk may be related to the cloudlets model
(e.g., Kuncic et al.\ 1997). If the standard cold disk exists on scales
larger than $10 r_g$, the disk should sample different solutions at
different radii (e.g., Chen et al.\ 1995). Then one might expect that the
inner, radiation dominated edge of the cold external disk puffs up at
large accretion rates, hiding at some epochs at least a part of the central
region and leading to separation of the external disk from the central hot
source. This may explain a delay of X-ray variations with respect to the
soft excess suggested by observations of some sources (e.g., Kaastra and
Barr 1989, NGC~5548). The inner region of the cold disk is dominated by
scattering, thus may be responsible for a variable soft excess continuum
(Magdziarz et al.\ 1998).

There is strong evidence that the reprocessing of X-rays is non-local,
at least in the best-observed Seyfert NGC~5548, since UV and optical
emission vary nearly precisely in phase (Krolik et al.\ 1991;  Clavel et
al.\ 1992). This strongly argues against stationary models of the
disk--patchy corona and a correlation between active regions has to be
postulated by, e.g., a much shorter time scale of energy transfer to the
active regions than the triggering time scale (Haardt et al.\ 1994). 
Alternatively, models with a central coherent X/$\gamma$ source can
naturally explain the global variability behavior. Recently, Magdziarz et
al.\ (1997) have detected in NGC~5548 a correlation between the total
brightness in X/$\gamma$-rays and the amount of reflection (i.e., the solid
angle of cold matter intercepting the X/$\gamma$ radiation in terms of
reprocessing models). In this picture the variability is driven by
geometrical changes of cold matter visible from the central X/$\gamma$
source on time scales consistent, as observed, with the thermal time scale on
the inner disk edge. In terms of a model proposed by Magdziarz and Blaes
(1997), an instability of the inner disk edge may lead to fragmentation
of the disk, thereby modulating the number of seed photons for
Comptonization in the central hot source. The observed non-linearity is
then related to the disk instability which produces a number of spatially
redistributed clouds or filaments of cold matter.  These could be
the reservoirs in the SOC model. Such a process may also produce signatures
of quasi-periodicity as a transient effect, which have been suggested in
some observations (Papadakis and Lawrence 1993, NGC~5548; Papadakis and
Lawrence 1995, NGC~4051).

Generally, the observed flux variability correlates with plasma parameters
inferred by the X-ray spectrum (cf.\ Haardt et al.\ 1997). Such a
correlation has been detected in a number of sources as a relation between
X-ray flux and spectral index in the sense that the source softens as it
brightens (e.g., Yaqoob and Warwick 1991, NGC~4151; Nandra et al.\ 1991,
NGC~5548; Papadakis and Lawrence 1995, NGC~4051; Leighly et al.\ 1997,
3C~390.9). If the variability is driven by changes of the cold matter
geometry, the increased number of seed photons would increase cooling of
the X/$\gamma$ source. This would make the X/$\gamma$ continuum soften
under the assumption that the plasma heating time scale is sufficiently long
(Magdziarz et al.\ 1998).  Such a mechanism, however, can not be universal,
at least in the case of MCG~-6-30-15, since the broad component of the
iron line in this object seems to be anti-correlated with the X/$\gamma$
continuum (Iwasawa et al.\ 1996). This should, generally, imply
anti-correlation of the reflection component, unless ionization of the
reprocessing matter is significantly variable.

\section{Conclusions}

We have discussed a new phenomenological idea of the broad-band spectral
formation and variability in Seyferts on the basis of disk models with
a transition region. This
region between the disk solutions forms an unstable and energetically
important zone of relatively cold ($\sim$ 100 eV) and fragmented matter.
This zone is presumably responsible for the soft excess continuum and
drives the variability related to the energy reprocessing by geometrical
changes of the phase which provides seed photons for Comptonization in the
central hot X/$\gamma$ region. If such a picture is true, then the soft
excess component may be the most important one for the energetics of the
source (cf.\ Magdziarz et al.\ 1998). However, the observational picture
of the soft excess is far from clear due to problems with spectral
deconvolution, and from the theoretical point of view physical conditions
in the transition region are complex and difficult to calculate (e.g.,
Igumenshchev et al.\ 1997). Strongly non-linear disk equations near the
transition region should simply allow generating a variety of Seyfert
types.

In the light of the above discussion instruments covering simultaneously
UV and soft X-ray bands, such as, e.g., SPECTRUM-X/$\gamma$, appear to be
the most important tool for understanding the physical nature of spectral
variability and energy reprocessing in the central source of AGNs.
 
We thank Omer Blaes,
Greg Madejski and Andrzej Zdziarski 
for useful discussion. 

\section{References}

\begin{enumerate}

\item
Chen, X., Abramowicz, M. A., Lasota, J.-P., Narayan, R. and Yi, I., ApJ,
{\bf 443}, L61 (1995).

\item
Haardt, F., Proc. Second Italian Workshop on Active Galactic Nuclei
(astro-ph/9612082) (1997). 

\item
Haardt, F., Maraschi, L. and Ghisellini, G., ApJ, {\bf 432}, L95 (1994).

\item
Haardt, F., Maraschi, L. and Ghisellini, G., ApJ, {\bf 467}, 620 (1997).

\item
Igumenshchev, I. V., Abramowicz, M. A. and Novikov, I., MNRAS, in press
(1997).

\item
Iwasawa, K., et al., MNRAS, {\bf 282}, 1038 (1996).

\item
Kaastra, J. S. and Barr, P., AA, {\bf 226}, 59 (1989).

\item
Kuncic, Z., Celotti, A. and Rees, M. J., MNRAS, {\bf 284}, 717 (1997).

\item
Leighly, K. M., Proc. of IAU Symp. 188, Kyoto, Japan, in press
(1997).

\item
Leighly, K. M., et al., ApJ, {\bf 483}, 767 (1997).

\item
Leighly, K. M. and O'Brien, P. T., ApJ, {\bf 481}, L15 (1997).

\item
Maciolek-Nied\'zwiecki, A., et al., in preparation (1998).

\item
Magdziarz, P., et al., in preparation (1998).

\item
Magdziarz, P., et al., Proc. 4th Compton Symposium,
in press (astro-ph/9707202) (1997).

\item 
Magdziarz, P. and Blaes, O., Proc. of IAU Symp. 188, Kyoto,
Japan, in press (astro-ph/9710183) (1997).

\item
Matt, G., Proc. Second Italian Workshop on Active Galactic Nuclei (1997).

\item
Mineshige, S., Ouchi, B. and Nishimori, H., PASJ, {\bf 46}, 97 (1994).

\item
Nandra, K., et al., MNRAS, {\bf 273}, 85 (1991).

\item
Papadakis, I. E. and Lawrence, A., MNRAS, {\bf 272}, 161 (1995).

\item
Papadakis, I. E. and Lawrence, A., Nature, {\bf 361}, 233 (1993).

\item
Turner, T. J., George, I. M., Nandra, K. and Mushotzky, R. F., ApJ, 
in press (1997).

\item
Yaqoob, T. and Warwick, R. S., MNRAS, {\bf 248}, 773 (1991).

\item
Zdziarski, A. A., et al., Proc. of the 2nd INTEGRAL Workshop, ESA
SP-382 (1997).

\end{enumerate}

\end{document}